\documentclass[a4paper,twocolumn,superscriptaddress,floatfix,showpacs,aps]{revtex4}
\usepackage[english]{babel}
\usepackage[utf8]{inputenc}
\usepackage{amsmath}
\usepackage{graphicx,epstopdf}
\usepackage{amssymb,epsfig,color,textcase}
\usepackage{hyperref}

\begin{document}

\title{Can the highly symmetric $SU(4)$ spin-orbital model be realized in $\alpha$-ZrCl$_3$?}

\author{A. V. Ushakov}
\affiliation{Institute of Metal Physics, Russian Academy of Science, S. Kovalevskaya Str. 18, 620041 Ekaterinburg, Russia}

\author{I. V. Solovyev}
\affiliation{Institute of Metal Physics, Russian Academy of Science, S. Kovalevskaya Str. 18, 620041 Ekaterinburg, Russia}
\affiliation{Ural Federal University, Mira Str. 19, 620002 Ekaterinburg, Russia}
\affiliation{International Center for Materials Nanoarchitectonics,
National Institute for Materials Science, 1-1 Namiki, Tsukuba,
Ibaraki 305-0044, Japan}

\author{S. V. Streltsov}
\email{streltsov.s@gmail.com}
\affiliation{Institute of Metal Physics, Russian Academy of Science, S. Kovalevskaya Str. 18, 620041 Ekaterinburg, Russia}
\affiliation{Ural Federal University, Mira Str. 19, 620002 Ekaterinburg, Russia}

\date{\today}

\begin{abstract}
We study physical properties of a potential candidate for the physical realization of the $SU(4)$ spin-orbital model - layered $\alpha$-ZrCl$_3$. Both DFT and DFT+U+SOC calculations show that this material most probably dimerizes at normal conditions. Therefore it is unlikely that symmetric $SU(4)$ model can be used for description of magnetic properties of $\alpha$-ZrCl$_3$ at low temperatures. In the dimerized structure electrons occupy molecular orbitals formed by the $xy$ orbitals. One might expect a non-magnetic ground state in this case. Interestingly the energy difference between various dimers packings is rather small and thus dimers may start to flow over the lattice as they do in Li$_2$RuO$_3$.
\end{abstract}

\maketitle

{\it Introduction.} Highly symmetric models play a special role not only in the condensed matter physics, but in a whole physics. The symmetry of such models of course manifests itself in their solutions, properties of the ground and excited states and finally in various observables. Therefore a lot of efforts were made in attempts to solve these models and also in search of their physical realizations. A special efforts were put into studying of highly symmetric spin and spin-orbital models, since they are especially important for description of magnetic materials. In particular it was shown that in case of the common-face geometry the Kugel-Khomskii spin-orbital Hamiltonian has unexpectedly high symmetry~\cite{Kugel2015,Khomskii2016}. Another example is the Kitaev model, which naturally appears in layered materials with the honeycomb lattice and heavy transition metal ions, such as Ir$^{4+}$ or Ru$^{3+}$. Strong spin-orbit coupling results in formation of effective moments $j_{eff}=1/2$, which in a specific common-edge geometry turns out to be coupled in a very anisotropic way~\cite{Jackeli2009}. One of the important results was a possibility of spin-liquid ground state realization in Kitaev materials. The quest for such a state in canonical materials Na$_2$IrO$_3$, Li$_2$IrO$_3$, and $\alpha$-RuCl$_3$ was unfortunately failed at normal conditions, see e.g. reviews~\cite{Winter2017,Takagi2019}, but further studies demonstrated that it can be indeed realized in external magnetic field~\cite{Johnson2015,Zheng2017}.

Meantime Yamada and co-authors~\cite{Yamada2018} noticed that other layered materials with the same structural motif - honeycomb lattices, but with one electron instead of one hole ($t_{2g}^1$ instead of $t_{2g}^5$) can be potentially described by a very symmetric spin-orbital Hamiltonian. They proposed that  $\alpha$-ZrCl$_3$ with one electron residing in the relativistic $j_{eff}=3/2$ manifold can be a physical realization of this model. In the present paper we performed {\it ab initio} study to check the hypothesis about realization of the $SU(4)$ symmetric spin-orbital model in this material.

\begin{figure*}[t!]
\includegraphics[width=4.8in]{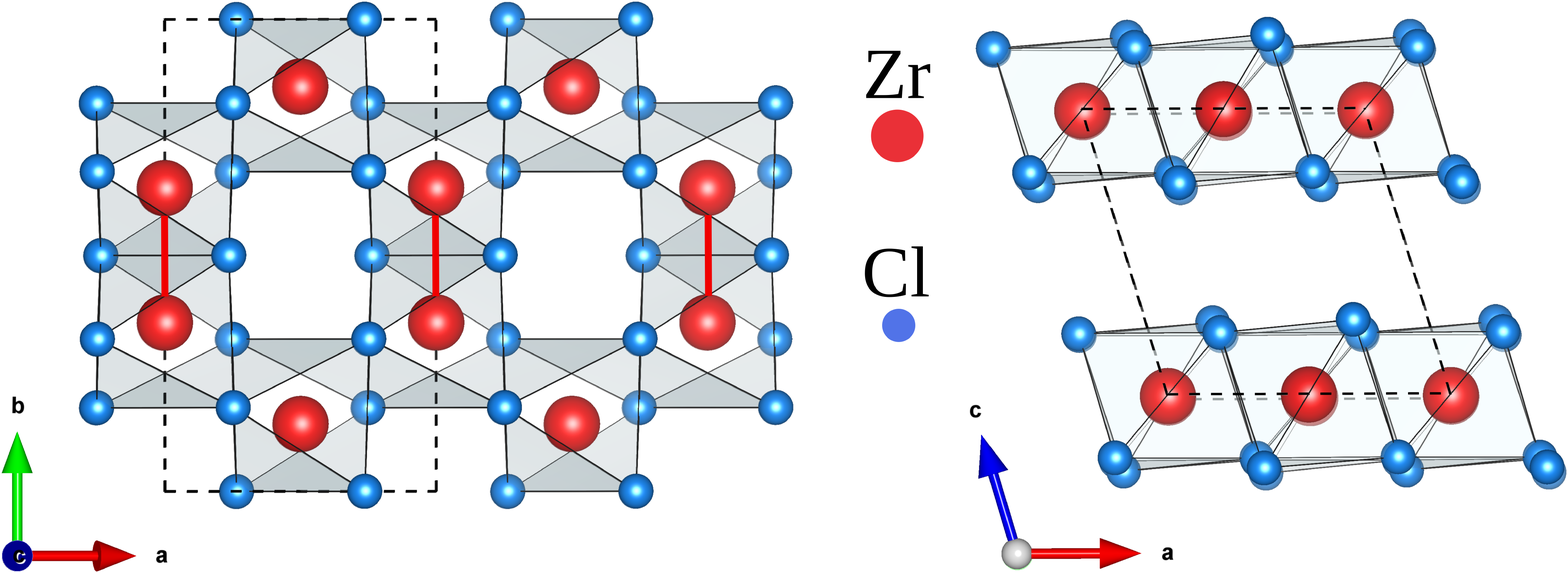} 
\caption{\label{fig:str} The crystal structure of $\alpha-$ZrCl$_3$ obtained by the optimization of $\alpha-$RuCl$_3$ structure. Zr--Zr dimers in honeycomb lattice are marked by red line.}
\end{figure*}

While the crystal structure of $\alpha$-ZrCl$_3$ has been previously experimentally refined, both small and main unit cell settings presented in Ref.~\cite{Swaroop1964} do not provide appropriate crystal structures of edge-sharing ZrCl$_6$ octahedra forming honeycomb layers (presented in left panel of Fig.~\ref{fig:str}). Therefore, the first important issue is establishing a probable candidate for the crystal structure of  $\alpha$-ZrCl$_3$. This can be done by DFT.

In order to test available computational tools we started with a sister material, $\alpha$-RuCl$_3$, which was thoroughly studied in the past decade. Initial structural model for $\alpha$-RuCl$_3$ was based on the P3$_1$12 space group~\cite{Stroganov1957}. However, more recent combined X-ray diffraction and DFT studies found the C2/m space group is more appropriate~\cite{Johnson2015}.

{\it Calculation details}. We used the generalized gradient approximation (GGA)~\cite{exchcorrp2} and projector augmented-wave (PAW) method as realized in the VASP code~\cite{vasp2} for the calculations.  We utilized hard Cl pseudopotential and considered $s$ states as valent ones for Zr and Ru. For GGA+U  calculations we used its version as presented in Ref.~\cite{Dudarev1998}. The cut-off energy was chosen to be 600 eV and the mesh of 6$\times$4$\times$4 points was used for integration over the Brillouin zone. We optimized cell volume, its shape and atomic positions in the structural relaxation, which was performed until the total energy change from one ionic iteration to another was larger than 0.1 meV.

{\it Results: structural optimization}.
It is interesting that without Coulomb correlations (treated on the level of GGA+U method) and the spin-orbit coupling  the lowest total energy for $\alpha$-RuCl$_3$ corresponds to the ferromagnetic order with nearly isotropic lattice, the distance between nearest in the honeycomb plane Ru ions is 3.48 \AA. For other magnetic structures in GGA we obtained two slightly different Ru-Ru bonds with bond-length difference of $\delta_{\rm Ru}\sim$0.06-0.16 \AA~(depending on the configuration). This is close to results of \cite{Johnson2015}, where $\delta_{\rm Ru}\sim$0.04 \AA.

Then, after testing the computational scheme, on $\alpha$-RuCl$_3$ one may perform the crystal structure optimization of $\alpha$-ZrCl$_3$. We used data of $\alpha$-RuCl$_3$ as a starting point and relaxed all possible parameters in magnetic GGA (cell volume, cell shape, atomic positions). The results are presented in Tab.~\ref{GGA-ZrCl3}.

Surprisingly the lowest in energy turned out to be not a uniform structure (all Zr-Zr bonds are the same), but the dimerized one with dimers being parallel to each other. It is important that Zr-Zr distance in this state is smaller than in Zr metal~\cite{Streltsov-2017}. 
The details of the electronic and magnetic structure as well as exchange interaction will be discussed further on, but already at this point one would expect that the dimerization is a result of formation strong molecular bonding (or orbital order) between two $t_{2g}$ orbitals looking towards each other. Stabilization of a single (per site) electron at the particular orbital will kill $SU(4)$ invariance of the spin-orbital model.  

It is worth noting that while $\alpha$-RuCl$_3$ has a structure with nearly regular hexagons (and this is exactly what was obtained for this compound in our calculations), it is known to dimerize under tiny pressure of  0.2 GPa~\cite{Bastien2018}. Furthermore, it dimerizes exactly in the same structure, featured by parallel orientation of the Ru-Ru dimers, as $\alpha$-ZrCl$_3$ in our calculations at the ambient pressure. In yet another material TiCl$_3$ with the same structural motif and the same (as in $\alpha$-ZrCl$_3$) $d^1$ configuration, a drop of magnetic susceptibility was  observed experimentally at $\approx$217~K~\cite{Ogawa1960}, which is likely associated with formation of the spin gap due to dimerization. Similar behaviour was observed in many other titanates~\cite{Isobe2002,Streltsov2008,Seidel2003,Schmidt2004}.

\begin{figure}[ht!]
\includegraphics[width=3.5in]{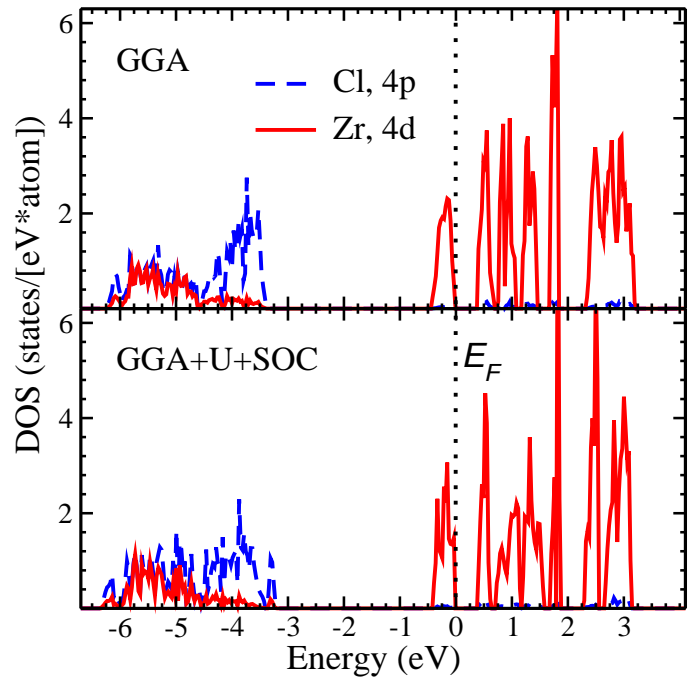} 
\caption{\label{fig:dos} The partial densities of states of $\alpha-$ZrCl$_3$ calculated in the GGA and GGA+U+SOC approximations for the dimerized structure with parallel dimers.}
\end{figure}

It is rather interesting that while the lowest in energy configuration corresponds to parallel dimers, the other one with armchair geometry is rather close in energy and one might expect that dimers might start to flow over the lattice in $\alpha$-ZrCl$_3$ at temperatures $\sim$ 500 K in the same way they do in Li$_2$RuO$_3$~\cite{Kimber2014}. This theoretical prediction would be very interesting to check experimentally.

Contrary to $\alpha$-RuCl$_3$, the lowest energy state of $\alpha$-ZrCl$_3$ is no longer ferromagnetic. This is interesting, because for metallic systems at the beginning of the band filling ($t_{2g}^1$ configuration) one could generally expect the ferromagnetic ground state~\cite{Heine}. However, $\alpha$-ZrCl$_3$ appears to be an insulator even at the GGA level (the band gap is about 0.3 eV). This will certainly break the phenomenological rule~\cite{Heine}, resulting in an AFM ground state.

{\it Results: electronic structure.}
The electronic structure of $\alpha$-ZrCl$_3$ in the dimerized phase is presented in Fig.~\ref{fig:dos}. It is rather similar to another dimerized transition metal compound with the same structural motif - Li$_2$RuO$_3$~\cite{Kimber2014,Pchelkina,Park2016}. These are Zr $4d$ states, which are in vicinity of the Fermi level. We chose the local coordinate system for any pair of ZrCl$_6$ octahedra in such a way that the axes are directed to Cl ions and the common edge is in the $xy$ plane. Then the $xy$ orbitals looking towards each other in this edge-sharing sharing geometry form molecular orbitals and this results in strong bonding-antibonding splitting seen in the density of states plot. Two electrons of the dimer occupy the bonding state and this results in the non-magnetic ground state. Other $t_{2g}$ states are  in between of these bonding and antibonding orbitals. The states at $\sim 3$ eV are $e_g$ orbitals of Zr.

\begin{figure}[ht!]
\includegraphics[width=2in]{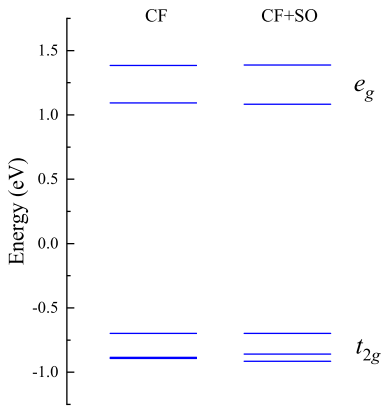} 
\caption{\label{fig:splitting} Splitting of Zr $4d$ levels caused by the crystal field (left) and simultaneously crystal field and spin-orbit interaction (right) in the optimized dimerized crystal structure.}
\end{figure}

In order to gain further insight we construct an effective five-orbital Hubbard-type model for Zr $4d$ bands using the Wannier functions technique and dimerized crystal structure, obtained from the optimization in the framework of GGA~\cite{review2008}. We start with the local, on-site effects - the crystal field splitting. Without spin-orbit coupling, the $t_{2g}$ levels are split by 8 and 186 meV, separating the lowest-middle and middle-highest levels, respectively (Fig.~\ref{fig:splitting}).  The spin-orbit coupling constant is about 70 meV. Quiet expectedly, it additionally splits the lowest $t_{2g}$ levels, but has little effect on the position of other atomic states. Thus, the crystal field, though not particularly strong, lifts the orbital degeneracy and substantially modifies the $j_{eff}=3/2$ character of the lowest energy states. 

Other very important parameters are the hopping integrals. One may compute averaged transfer integrals connecting occupied ($o$) states and occupied and unoccupied ($u$) states of the nearest sites $i$ and $j$: $t_{ij}^{oo}$ and $t_{ij}^{ou} = \left( \sum_{b \ne o} t^{ob}_{ij}t^{bo}_{ji} \right)^{1/2}$, respectively, where $b$ is the orbital index. Then, for the dimerized structure $t_{ij}^{oo} = -1.262$ eV  and $t_{ij}^{ou} = 0.136$ eV. A very large  hopping  between occupied orbitals results in bonding-antibonding splitting $\sim$ 2.5 eV. 

For the uniform structure we have $t_{ij}^{oo} = 0.067$ and $t_{ij}^{ou} = 0.100$ eV. Then one can calculate exchange parameters of the Heisenberg model using superexchange theory for this structure.
In the simplest approximation, neglecting asphericity of on-site Coulomb ($U$) and exchange ($J_H$) interactions in the five-orbital model, $t_{ij}^{oo}$ contributes to the AFM coupling as $-\frac{(t_{ij}^{oo})^2}{U}$, while $t_{ij}^{ou}$ contributes to both FM and AFM coupling as $-\frac{(t_{ij}^{ou})^2}{U}$ and $\frac{(t_{ij}^{ou})^2}{U-J_H}$, respectively. Using realistic estimates for $U$ and $J_H$ (see below), this yields weak total exchange coupling $J \sim 0.26$ meV.

{\it Results: correlation effects.} It has to be stressed that while all results presented above were obtained without account of on-site Coulomb correlations and the spin-orbit coupling, their inclusion to the calculation scheme does not change the main conclusion that $\alpha$-ZrCl$_3$ tends to dimerize. 

In order to have realistic estimation of Hubbard $U$ and Hund's $J_H$ parameters we used the constrained random phase approximation (cRPA)~\cite{review2008}, which yields $U=1.53$ and $J_H=0.58$ eV. Particularly, the Coulomb $U$ is strongly screened, as expected at the beginning of the band filling~\cite{review2008}. These values were used in the subsequent GGA+U+SOC calculations. Basically $U$ renormalizes GGA energy differences between different solutions discussed above, but it does not change the ground state structure. The energy of the armchair configuration is 68 meV and of the AFM zigzag is 28 meV. Zr-Zr bond lengths in parallel dimers configuration are 3.094~\AA. 

One also needs to comment on the importance of the spin-orbit coupling. The orbital moment in the lowest in energy structure is tiny ($\sim 10^{-3}\mu_B$) and therefore one may neglect this interaction in dimerized structure. This is rather natural since dimerization results in a strong deformation of the octahedra, $t_{2g}$ manyfold is split and the orbital moment gets quenched. Formation of molecular orbitals only helps this quenching. We note, however, that in some dimerized or trimerized structures the spin-orbit may play some role~\cite{Terzic2015,Streltsov2017,Komleva2020}.

\begin{table}[t]
\centering
\caption{Total energies (per f.u.), Zr-Zr bond lengths for various magnetic and structural configurations. Results of magnetic GGA calculations. Note, that Neel AFM structure converges to the nonmagnetic (NM), while stripe AFM to the parallel dimers solution.}
\begin{tabular}{l c c}
\hline
\hline
Configuration & Energy   & $d($Zr-Zr$)$ \\
\hline 
NM uniform        &  220 meV & 3.581$\times$3 \AA        \\ 
FM uniform        &  176 meV & 3.607, 3.616$\times$2 \AA \\ 
AFM zigzag        &  107 meV & 3.952, 3.435$\times$2 \AA \\ 
Armchair dimers   &  50 meV & 3.068, 3.972$\times$2 \AA   \\
Parallel dimers   &  0  & 3.089, 3.928, 3.937 \AA   \\
\hline
\hline
\end{tabular}
\label{GGA-ZrCl3}
\end{table}

{\it Conclusions.}
Using results of first principles electronic structure calculations, we have argued that the ideal honeycomb structure of $\alpha$-ZrCl$_3$ has a strong tendency towards symmetry lowering via the formation of the Zr-Zr dimers, similar to other transition-metal compounds having formal electronic configuration $d^{1}$. Although the dimerization scenario has many interesting aspects, it seems to be at odds with the realization in this material high $SU(4)$ symmetry required for the formation of the spin-orbital liquid state~\cite{Yamada2018}. 

{\it Acknowledgements.}
S. Streltsov. is grateful to A. Gubkin for help with analysis of structural data in Ref.~\cite{Swaroop1964}, to H. Jeschke for consultations on the crystal optimization of $\alpha$-RuCl$_3$ and especially to G. Jackeli and D.I. Khomskii for various stimulating discussions on spin-orbital physics.

This work was supported by the Russian Science Foundation through RSF 20-62-46047 research grant.

\end{document}